\documentclass{article}

\PassOptionsToPackage{numbers, compress}{natbib}

\usepackage[dblblindworkshop, final]{neurips_2025}
\workshoptitle{Multimodal Representation Learning for Healthcare}

\usepackage{subcaption}  
\usepackage{amsmath}
\usepackage{authblk}
\usepackage{setspace}
\usepackage{hyperref}
\usepackage[utf8]{inputenc} 
\usepackage[T1]{fontenc}    
\usepackage{hyperref}       
\usepackage{url}            
\usepackage{booktabs}       
\usepackage{amsfonts}       
\usepackage{nicefrac}       
\usepackage{microtype}      
\usepackage{xcolor}         
\usepackage{graphicx} 
\usepackage{comment}      
\usepackage{multirow}

\title{A learning health system in Neurorehabilitation as a foundation for multimodal patient representation}

\author[1,2]{Thomas Weikert}
\author[3,4]{Eljas Roellin}
\author[3]{Lukas Heumos}
\author[3,4]{Fabian J. Theis}
\author[2]{Diego Paez-Granados}
\author[1]{Chris Easthope Awai}

\affil[ ]{\texttt{\{thomas.weikert, chris.awai\}@llui.org}}
\affil[ ]{\texttt{diego.paez@hest.ethz.ch}}
\affil[ ]{\texttt{\{eljas.roellin, lukas.heumos, fabian.theis\}@helmholtz-muenchen.de}}
\affil[ ]{\vspace{0.8em}}

\affil[1]{Data Analytics \& Rehabilitation Technology (DART) Lab, Lake Lucerne Institute, Switzerland}
\affil[2]{Spinal Cord and Artificial Intelligence (SCAI) Lab, ETH Zurich and Swiss Paraplegic Research, Switzerland}
\affil[3]{Institute of Computational Biology, Helmholtz Center Munich, Germany}
\affil[4]{Department of Mathematics, Technical University of Munich, Germany}

\begin{document}

\maketitle

\begin{abstract}
Neurological disorders represent a growing global health burden requiring long-term, interdisciplinary rehabilitation. Computational neurorehabilitation (compNR) - the use of data-driven and model-based approaches to personalize treatment - offers new opportunities for precision rehabilitation. However, its clinical deployment is limited by fragmented data systems, poor interoperability, and low clinician engagement in model development. We embed the learning health system (LHS) framework in Neurorehabilitation through integration of multimodal data collection, model computation, and clinical visualization that enables clinician–ML collaboration in everyday neurorehabilitation practice. The system facilitates structured digital data capture, secure computational processing, and interoperable visualization of patient trajectories. Through a real-world deployment in stroke rehabilitation, we demonstrate how such an infrastructure bridges the gap between research models and clinical use, showcasing one approach to a translational pathway for compNR.
\end{abstract}

\section{Introduction}
The World Health Organization indicates that one in three people live with a health condition that could benefit from rehabilitation \cite{gimigliano2017rehabilitation}, with numbers projected to rise significantly as populations age and stroke incidence outpaces other non-communicable diseases \cite{devaux2020will, chong2024global, cheng2024projections}. At the same time, there is a growing shortage of skilled workers and a high burnout rate among healthcare professionals \cite{de2020burnout}, raising pressing questions about future financing for rehabilitation. To address these issues, there is a critical need to optimize rehabilitation in terms of delivery efficacy, cost-effectiveness, and outcome prediction. The application of computational neurorehabilitation (compNR) holds the potential to transform rehabilitation services by achieving more with fewer resources\cite{panch2018artificial}. CompNR seeks to optimize treatment by modeling neural plasticity and learning dynamics to predict the most effective interventions for each patient \cite{reinkensmeyer2016computational, johnson2022computational}. Yet, embedding these methods into clinical workflows and distilling clinically actionable and meaningful insights from them demands collaborative frameworks that connect clinicians and ML researchers.
Recent work demonstrates that reinforcement learning \cite{ye2025ai}, causal inference \cite{cotton2024causal}, and digital twin modeling \cite{katsoulakis2024digital} can identify patient-specific therapeutic strategies. Wearable and ambient sensors provide continuous, high-frequency physiological and behavioral data relevant for adaptive modeling \cite{adans2021enabling, pohl2025construct, lohse2024validation}, tracking stroke-specific movement patterns \cite{pohl2022accuracy, pohl2022classification} and enabling novel biomechanical metrics \cite{uhlrich2023opencap, unger2024differentiable}.  There are already examples of successful individual health trajectory prediction \cite{moen2025towards, shmatko2025learning}, however, the data required for these models is only available from detailled research studies and is not collected in clinical routine. Especially for compNR that seeks to inform on each individual treatment session, a large and more data-dense foundation - both in terms of temporal resolution and detail, is indispensable.  Low data interoperability, siloed infrastructures across clinical and home contexts, and inadequate provisions for continuous sensor streams in existing frameworks \cite{brasier2020device, 9762855} prevent the creation of data at scale. A solution that allows the sourcing of this data directly from clinical routine promises larger and more diverse and equitable datasets, which can be used to train high performing, robust and trustworthy prediction systems.
To address this gap, we propose an embedded learning health system (LHS) \cite{menear2019framework, steel2025learning} that connects data acquisition, computational modeling, and data exploration \cite{heumos2024open} into a closed loop between patients, clinicians, and ML models. This work focuses on the translational aspects of deploying such systems in clinical neurorehabilitation environments and fostering sustained clinician–ML collaboration through modular, interoperable infrastructure that generates value for many stakeholders in the system: patient, clinicians, and researchers.

\section{Method}
We developed a LHS embedded within clinical practice to enable compNR. The design followed an iterative, user-centered approach grounded in contemporary LHS frameworks \cite{menear2019framework, steel2025learning}. Our taxonomy encompasses three categories: (1) a technology platform, (2) a rapid cycle model for prototyping, and (3) a multidisciplinary team and leadership. Together these elements create a closed learning loop in which multimodal clinical data are transformed into actionable insights that feed back into care delivery through clinical staff.

\begin{figure*}[htbp]
    \centering
    \begin{subfigure}[t]{0.48\textwidth}
        \centering
        \includegraphics[width=\textwidth]{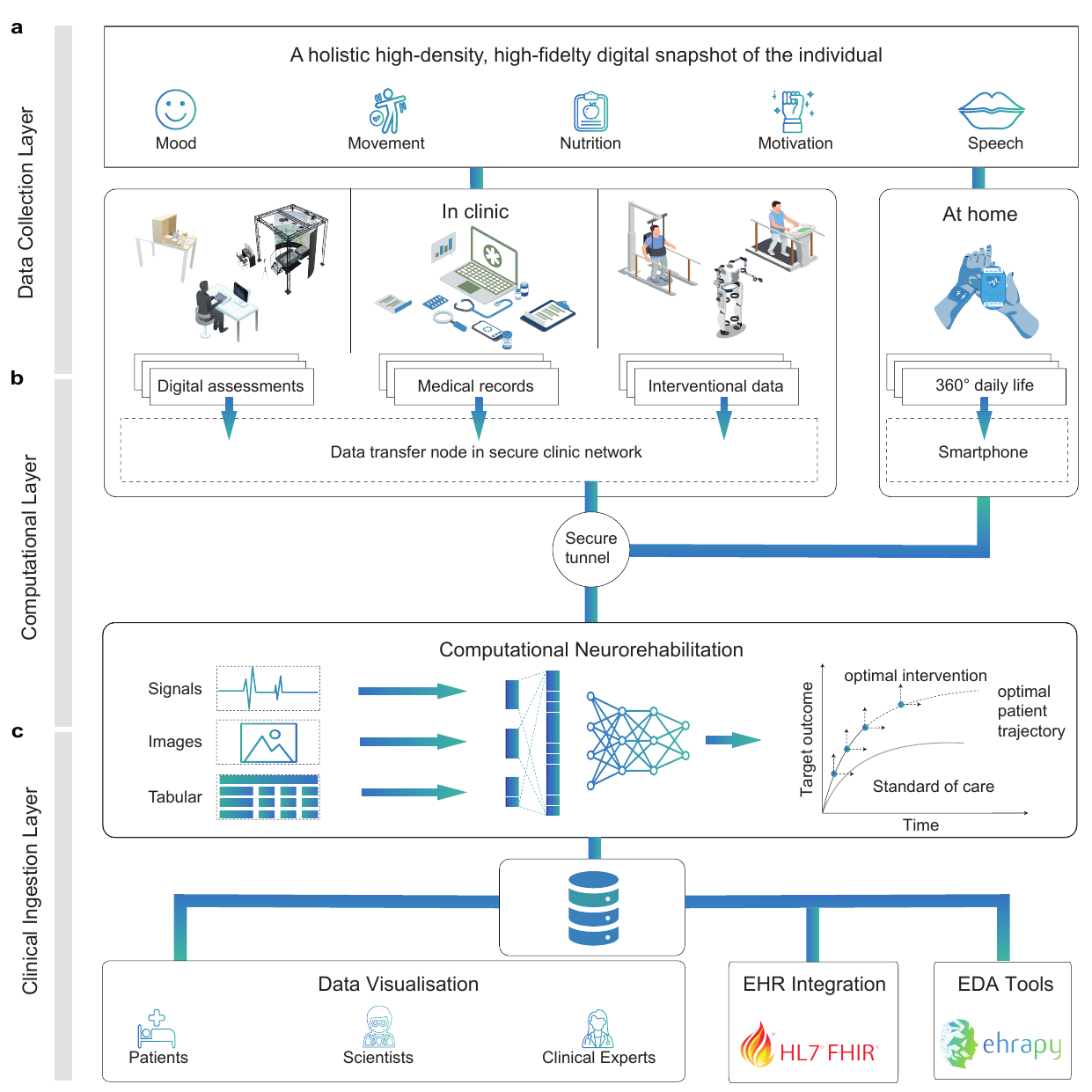}
        \caption{}
        \label{fig:HC-setup-a}
    \end{subfigure}
    \hfill
    \begin{subfigure}[t]{0.48\textwidth}
        \centering
        \includegraphics[width=\textwidth]{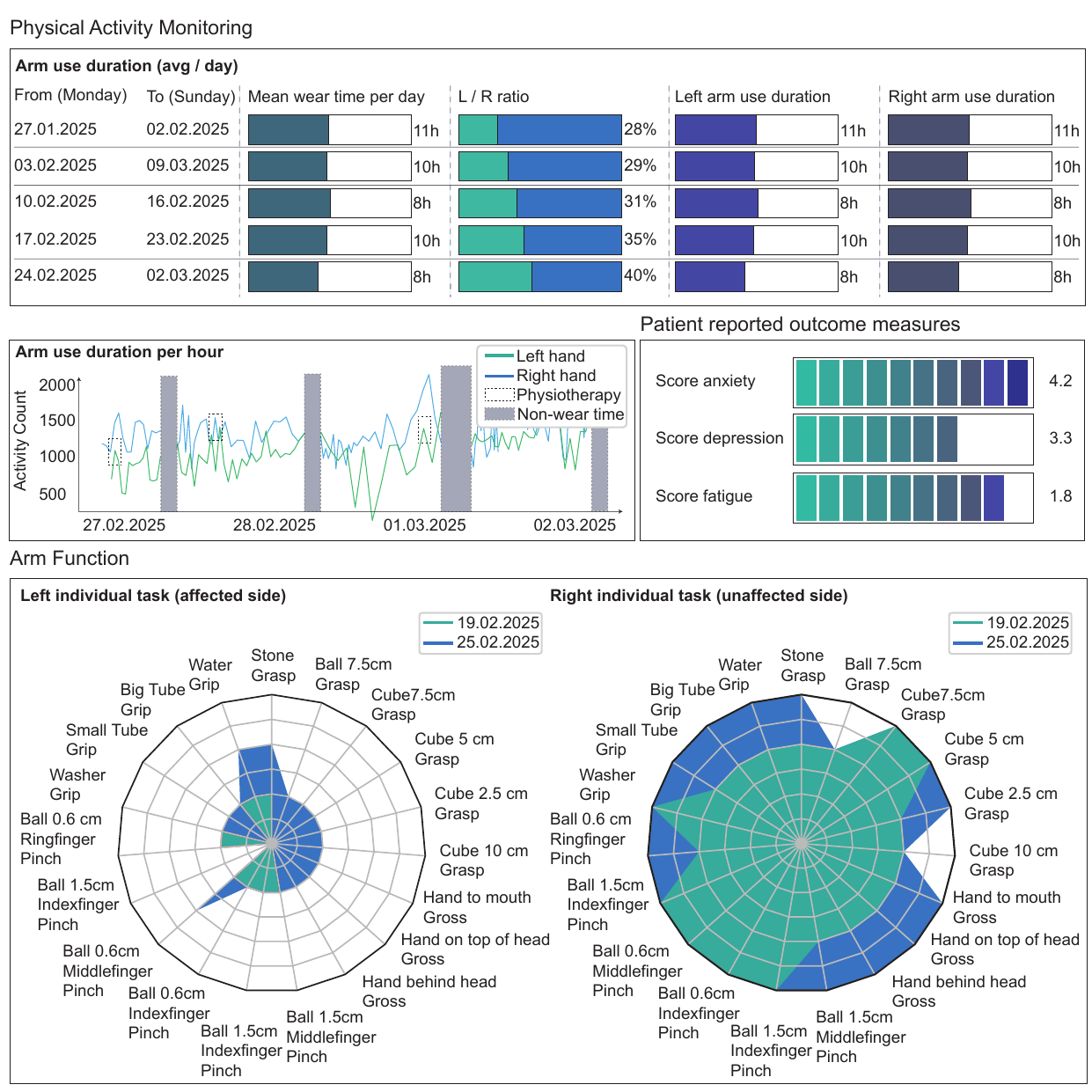}
        \caption{}
        \label{fig:HC-setup-b}
    \end{subfigure}
    
    \vspace{0.5cm} 
    
    \begin{subfigure}[t]{\textwidth}
        \centering
        \includegraphics[width=0.50\textwidth]{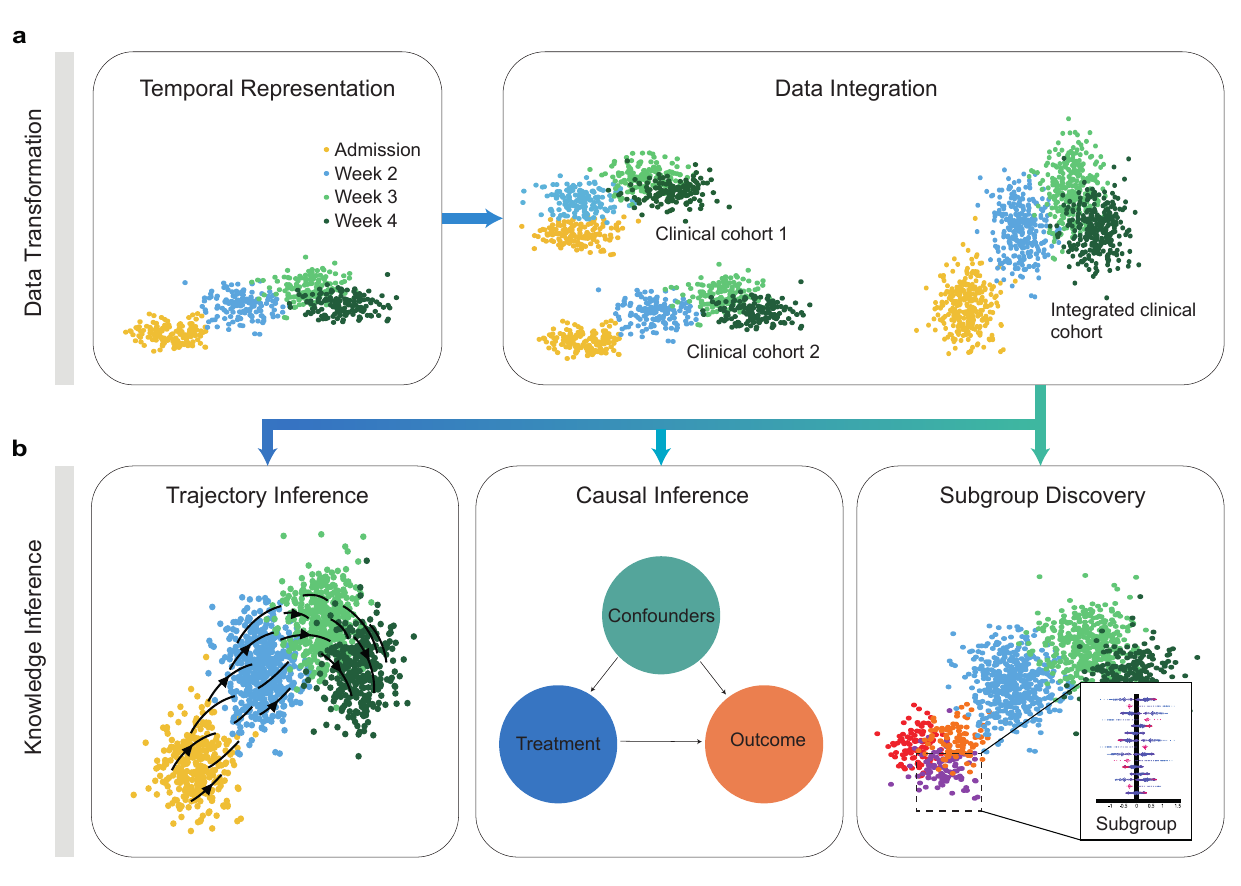}
        \caption{}
        \label{fig:HC-setup-c}
    \end{subfigure}
    
    \caption{(a) Technology platform: Schematic overview depicting three layers. Data collection from clinical and home sources. Computational pipeline from data to optimal intervention prediction. Clinical visualization and EHR integration. (b) Dashboard visualization. (c) Data exploration. Multi-modal rehabilitation data from digital assessments, medical records, and interventions across clinical cohorts enable downstream analyses including causal inference, trajectory modeling, and subpopulation discovery.}
    \label{fig:HC-setup}
\end{figure*}

Our technology platform (Figure \ref{fig:HC-setup}) is embedded within a member clinic of a national neurorehabilitation consortium and specifically tailored for stroke care. Data collection tools, usable both in-clinic and at home, are built using off-the-shelf computers, smartphones, or tablets, with applications developed in Android Studio or similar environments. Additionally, traditional data silos - such as robotics, imaging, and gait- and research laboratories are integrated through automated or semi-automated file transfer protocols \cite{pena2024ros}. This is achieved by providing a modular and extensible data collection application infrastructure at level of the clinic. The platform also interfaces with the clinic’s electronic health record (EHR) system, enabling weekly automated HL7 data extraction to our on-premises data encryption node that relays all data the the compute core. Equally, we can return data daily to the clinic's EHR system.

At the core of our learning health system (LHS) is a rapid-cycle model for prototype development of digital data capture tools. These tools enable enhanced data capture by providing quality guidance, directly transferring data to electronic health records, and offering opportunity for instrumentation. For each tool, we began with an existing standardized assessment battery covering movement, speech, and patient-reported outcomes, and enhanced it with wearable movement sensors, voice, and video recordings where appropriate. Our goal was to support clinical practitioners in manually evaluating patients’ health status across traditional rehabilitation domains. Through interviews, we identified clinician needs and operational constraints. Based on this, our development priorities were low capture time, high usability of digital tools, and clinical interpretability. Following an initial development phase, each of the proposed solutions is tested in a limited pilot, during which user feedback is continuously collected. This cycle of development, testing, and refinement is repeated until convergence - typically after 1-3 iterations. Using this process, we assembled an enhanced assessment battery of >10 different data collection tools that are fully integrated with patients’ medical records.

The backbone of our LHS is a multidisciplinary team comprising clinicians, data scientists, rehabilitation engineers, and leadership stakeholders. The team is structured around specific assessments. Each assessment team includes a biomedical engineer responsible for developing methodology, a software engineer who ensures platform connectivity and scalability, a life scientist who oversees usability and requirement engineering, and a clinician who provides feedback and patient perspectives. These team members collaborate continuously through workshops and regular update meetings. Leadership further ensures that the system’s evolution aligns with broader organizational and translational objectives, supporting both clinical advancement and research innovation.

\section{Result}

The LHS has been deployed continuously in clinical practice for two years, capturing over 2,500 patient-days of Physical Activity Monitoring with IMU sensors \cite{pohl2025construct} data to date. The monitoring system tracks activity levels and bilateral arm use between 8 to 14 hours per day, requiring only 15-20 minutes of setup time per patient. This approach provides longitudinal insights into rehabilitation trajectories, enabling clinicians to assess patient progress during therapy sessions and monitor real-world functional recovery throughout the day. We also recorded 230 digitally enhanced Action Research Arm Tests (ARAT) using IMU sensors and video recordings \cite{weikert2025automated, homm2025identiarat, unger2024upper}. The data collection and dashboard tools iteratively developed demonstrate high usability and high user satisfaction \cite{unger2025usability}. For upper-limb assessments, ratings were strong (System Usability Scale (SUS): 84 ± 5) with an additional measurement time of approximately 120 seconds and a training requirement of 2-3 supervised sessions. Lower limb assessments reflected this with similar satisfaction levels (SUS: 86 ± 4) and less than 60 seconds of added time per measurement.

Clinicians reported reduced documentation effort, improved access to longitudinal and complex information, and greater engagement with detailed data insights. The compute pipeline consistently decrypted, processed, and reintegrated data into clinical systems and analytic endpoints through automated orchestration, achieving a success rate above 90\%.

\begin{table*}[h!]
    \centering
    \caption{Automated assessment battery embedded into clinical practice: influx rates in a small rehabilitation clinic}
    \label{tab:clinical_outcomes} 
    \renewcommand{\arraystretch}{1.5} 
    \normalsize
    \resizebox{\textwidth}{!}{%
    \begin{tabular}{p{3cm} | l | p{6.cm} | p{2.5cm} | p{2cm} | p{2cm}}
        \hline
        \textbf{Clinical Outcome} & \textbf{Test} & \textbf{Measure} & \textbf{Traditional Scoring} & \textbf{Assessments per week} & \textbf{Wearable Sensors} \\
        \hline
        \multirow{1}{*}{Arm Function}  & Action Research Arm Test & Assesses grasp, grip, pinch, and gross motor movements & Trained therapist / pen \& paper & 2x & IMU, Video \\
        \hline
        \multirow{1}{*}{Activity Levels} & Physical Activity Monitoring & Quantifies physical activities in low, moderate, and vigorous times & Trained therapist & 3x & IMU \\
        \hline
        \multirow{1}{*}{Arm Use} & Physical Activity Monitoring & Quantifies the use of left and right arm within and outside of therapy & Trained therapist & 3x & IMU \\
        \hline
        \multirow{1}{*}{Linguistic Skills} & Frenchay Dysarthria Assessment (FDA) & Evaluates disorders resulting from neuromuscular impairments  & Trained therapist / pen \& paper & 3x & Tablet, Audio \\
        \hline
        \multirow{1}{*}{Linguistic Skills} & Bogenhausener Dysarthrieskalen (BoDyS) & Evaluates dysarthria in adults with neurological condition  & Trained therapist / pen \& paper & 3x & Tablet, Audio \\
        \hline
        \multirow{1}{*}{Walking Speed} & 10m Walking Test & Measures walking speed over a fixed distance (m/s) & Stopwatch & 3x & Video \\
        \hline
        \multirow{1}{*}{Perceived Fatigue} & Fatigue Severity Scale & Reports impact of fatigue on daily function & Pen \& paper & 5x & Tablet \\
        \hline
        \multirow{1}{*}{Perceived Anxiety} & Hospital Anxiety \& Depression Scale & Reports anxiety and depression in hospitalized patients and outpatient settings & Pen \& paper & 5x & Tablet \\
        \hline
        \multirow{1}{*}{Perceived Depression} & Beck Depression Inventory II Scale & Reports the severity of depressive symptoms & Pen \& paper & 5x & Tablet \\
        \hline
        \multirow{1}{*}{Perceived Sleep} & Epworth Sleepiness Scale & Reports daytime sleepiness and likelihood of falling asleep & Pen \& paper & 5x & Tablet \\
        \hline
        \multirow{1}{*}{Perceived Fatigue} & Fatigue Scale for Motor and Cognitive Functions & Reports motor and cognitive aspects of fatigue for MS patients & Pen \& paper & 5x & Tablet \\
        \hline
    \end{tabular}
    \label{tab:assessment_battery} 

    }
\end{table*}

\section{Discussion}
This work illustrates how embedding a LHS within clinical workflows can transform data collection from a research task into a practical clinical resource. By designing around clinician needs - low capture time, integration with existing assessments, and meaningful visualization, we achieved high user satisfaction and long-term adoption.
From a translational perspective, the embedded LHS provides a blueprint for clinician - ML co-development: researchers gain continuous multimodal datasets and real-world feedback, while clinicians benefit from interpretable, data-driven decision support.

We currently work on extending the digital assessment battery to include domains such as sleep and nutrition. Future work will extend this collaboration toward explainable, temporal modeling of rehabilitation trajectories. By learning representations that integrate multimodal data - movement, speech, physiological signals, and subjective reports—models can identify subpopulations, estimate treatment response windows, and generate causal insights into recovery dynamics. Clinician interaction with these models will be key for iterative validation, trust calibration, and refinement of intervention strategies.

\section{Conclusion}

We demonstrate a clinically embedded, modular i-health system enabling the structured collection of data, multimodal data integration across disparate sources, computational neurorehabilitation models, and clinician-centered visualization. Through sustained collaboration between ML researchers and healthcare professionals, this system bridges the gap between algorithmic development and clinical use.
Our deployment shows that translational pathways for multimodal AI in healthcare require more than technical performance—they depend on co-design, interoperability, and trust. The framework presented here provides a scalable model for embedding computational methods into everyday clinical decision-making, paving the way toward adaptive, data-driven neurorehabilitation.

\bibliographystyle{unsrtnat}
\bibliography{references}

\end{document}